\documentclass[aps,prl,twocolumn,groupedaddress,showpacs]{revtex4}

\usepackage[dvips]{graphicx,color}

\begin{document}

\title{Apparent Clustering of Intermediate-redshift Galaxies as a
Probe of Dark Energy }

\author{Takahiko Matsubara}
\email[]{taka@a.phys.nagoya-u.ac.jp}
\affiliation{Department of Physics and Astrophysics,
Nagoya University, Chikusa, Nagoya 464-8602, Japan}
\author{Alexander S. Szalay}
\email[]{szalay@jhu.edu}
\affiliation{Department of Physics and Astronomy, 
The Johns Hopkins University, Baltimore, MD 21218}

\date{\today}

\begin{abstract}
We show the apparent redshift-space clustering of galaxies in redshift
range of 0.2--0.4 provides surprisingly useful constraints on dark
energy component in the universe, because of the right balance between
the density of objects and the survey depth. We apply Fisher matrix
analysis to the the Luminous Red Galaxies (LRGs) in the Sloan Digital
Sky Survey (SDSS), as a concrete example. Possible degeneracies in the
evolution of the equation of state (EOS) and the other cosmological
parameters are clarified.
\end{abstract}

\pacs{98.80.Es, 95.35.+d, 98.62.Py}

\maketitle

The dark energy, such as the cosmological constant, has now turned out
to be a necessary element to understand our universe. There are many
indirect suggestions for the dark energy, including the age of the
universe, the formation of the large-scale structure, the number count
of the galaxies, and so on \cite{car92}. More striking evidences stem
from the combination of acoustic peaks of temperature fluctuations in
cosmic microwave background radiation (CMB) \cite{bal00} and the
Hubble diagram of the type Ia supernova \cite{rie98}. Natural
expectation for the dark energy is arisen from the vacuum fluctuations
of quantum fields, although the smallness of its observed value is
extremely unnatural \cite{wei89}.

Since the energy density of the cosmological constant is constant in
time by definition, one needs an extremely suspicious fine-tuning of
120 digits to set a correct value at an initial stage (probably Planck
time) so that the energy density of the cosmological constant should
be comparable with the matter density today. Alternatively, such
stringent fine-tuning should be moderated if the dark energy component
consists of some dynamical field such as ``quintessence''
\cite{rat88,cal98}. In general, such dynamical dark energy affects
cosmological observations through a time-dependent EOS, which is
characterized by a parameter $w(t) = p/\rho$. If $w$ is found to be
different from $-1$, the dark energy is proven to be different from the
simple cosmological constant. Current observational data are
consistent with the cosmological constant \cite{bea02,dor02}, although
they are still not enough to impose sufficient constraints on the EOS.

Since the dark energy has profound implications on the nature of the
universe, it is of great importance to explore the origin of dark
energy. Near-future observations will enable us to clarify whether the
dark energy is a mere cosmological constant or something else.
Cosmological observations, such as the Type Ia supernovae
\cite{tur97}, CMB fluctuations \cite{cal98,dor02,bac02}, cluster mass
function \cite{hai01}, weak lensing field \cite{hui99}, and so on, are
useful tools to probe the nature of the dark energy.

It is shown \cite{mat96} that an application of the Alcock-Paczynski
(AP) test \cite{alc79} to the redshift-space correlation function of
the high-redshift objects can be a useful probe of the cosmological
constant. In literatures, this method is mainly applied to the
Lyman-$\alpha$ forest \cite{hui99a}, the Lyman-break galaxies
\cite{nai99}, and the quasars \cite{pop98}, etc.

A drawback of the high-redshift objects is their sparseness.
Relatively high shot noise prevents to accurately determine the
correlation function. On the other hand, a typical sample of normal
galaxies is dense enough to have sufficiently small shot noise, while
it is too shallow to apply the AP test. Instead, a sample of
intermediate-galaxies, such as the LRGs \cite{eis01} in the SDSS has a
right balance between density and depth \cite{mat01, mat02}.

Therefore, a natural expectation is that the clustering of
intermediate-galaxies can distinguish various dark energy models quite
well. In fact, the AP test around redshift 0.5 is suggested to be
promising \cite{kuj02}. The purpose of this Letter is to
quantitatively investigate what constraints can be imposed on the
parameter space of dark energy models from intermediate-redshift
galaxies and to clarify possible degeneracies with other cosmological
parameters.

The clustering of objects in redshift space is characterized by the
correlation function. In linear regime, the full information on the
clustering is contained in the two-point correlation function, if the
non-Gaussianity of the density field is negligible. In a homogeneous,
isotropic space, the two-point correlation function $\xi(r)$ is a
function of only a separation $r$ of the two-points. However, the
redshift-space clustering in reality is neither homogeneous nor
isotropic. Peculiar velocities, evolutionary effects, and cosmological
geometry introduce inhomogeneity and anisotropy in the observed space.
Therefore, the two-point correlation function in apparent redshift
space is generally expressed as $\xi(z_1,z_2,\theta)$, where $z_1$,
$z_2$ are the redshifts of the two-points and $\theta$ is the apparent
angle between two-points from the observer. The full analytic
expression for the apparent correlation function in linear regime is
known \cite{mat00}.

It is not trivial how one can analyze the three dimensional function
$\xi(z_1,z_2,\theta)$. Instead of directly evaluate the two-point
correlation function, we have been developing a likelihood method for
the density field \cite{vog96,mat00a,mat01}, which directly analyze
the probability of the observed density field given a cosmological
model. A fast algorithm to compute the correlation matrix is crucial
in this analysis. A detailed description of one of our algorithms is
given in \cite{mat02}, in which distant-observer approximation is
assumed. We generalize this algorithm to the one without this
approximation, using the analytic expression of \cite{mat00}. Apart
from this generalization, and from inclusion of the dark energy
degrees of freedom, the algorithm we employ in this Letter is similar
to that in \cite{mat02}, the details of which will be published
elsewhere.

Here we briefly summarize major points of the generalization. In our
likelihood analysis of the redshift-space clustering, first we place
smoothing cells in redshift space and then count the number of object
$n_i$ in each cell $i$. When the non-Gaussianity of the distribution
is negligible, the correlation matrix $C_{ij} = \langle n_i n_j
\rangle/(\langle n_i \rangle\langle n_j \rangle) - 1$ fully
characterizes the statistical property of the clustering. The
correlation matrix is given by numerical integrations of the analytic
form of $\xi(z_1,z_2,\theta)$ which is generalized to take into
account nontrivial the EOS for dark energy component. Both
cosmological distortions and peculiar velocity distortions are taken
into account in this analytic form. A spherical cell approximation
\cite{mat02} greatly simplifies this integration without implementing
the numerical integration cell by cell. Then shot noise term is added
to the correlation matrix.

Most of the modification of the analytic form of the correlation
function is originated from the non-standard time-dependence of the
Hubble parameter. When the dark energy has a non-trivial EOS as a
function of redshift, $w(z)$, the time-dependent Hubble parameter is
given by
\begin{eqnarray}
   H(z) &=& H_0
   \biggl[
      (1+z)^3 \Omega_{\rm M0} -
      (1+z)^2 \Omega_{\rm K0}
   \biggr.
\nonumber\\
&&\quad
   \left.
      +\; \exp\left(3\int_0^z \frac{1 + w(z)}{1+z} dz\right)
      \Omega_{\rm Q0}
   \right]^{1/2},
\label{eq1}
\end{eqnarray}
where $\Omega_{\rm M0}$, $\Omega_{\rm Q0}$ are the present density
parameter of matter and dark energy components, respectively, and
$\Omega_{\rm K0} = \Omega_{\rm M0} + \Omega_{\rm Q0} - 1$ is the
present curvature parameter. Evaluation of the analytic form of the
apparent correlation function requires the comoving distance
$\chi(z)$, the growth factor $D(z)$, and the logarithmic derivative of
the growth factor $f(z) = d\ln D/d\ln a$ as functions of redshift. All
these quantities are determined by the time-dependent Hubble
parameter. The comoving distance is simply given by $\chi(z) =
\int_0^z H^{-1}(z)dz$. The growth factor is the growing solution of
the differential equation, $\ddot{D} + 2H\dot{D} - \frac32 {H_0}^2
\Omega_{\rm M0} (1+z)^3 D = 0$, where dot represents the
differentiation with respect to the time $t$. It is useful to rewrite
this equation into the following set of equations,
\begin{eqnarray}
&&\!\!\!\!\!\!\!\!\!\!\!\!
   \frac{d\ln D}{d\ln a} = f,
\label{eq3a}\\
&&\!\!\!\!\!\!\!\!\!\!\!\!
   \frac{df}{d\ln a} = -f^2
   - \left(1 - \frac{\Omega_{\rm M}}{2}
      - \frac{1+3w}{2} \Omega_{\rm Q}
     \right) f
   +\; \frac32 \Omega_{\rm M},
\label{eq3b}
\end{eqnarray}
where $a=(1+z)^{-1}$ is the scale factor of the universe, and
$\Omega_{\rm M}(z) = {H_0}^2 (1+z)^3 \Omega_{\rm M0}/H^2(z)$,
$\Omega_{\rm Q}(z) = {H_0}^2 \exp(3\int_0^z \frac{1+w}{1+z} dz)
\Omega_{\rm Q0}/H^2(z)$ are the time-dependent density parameters of
matter and dark energy, respectively. The Runge-Kutta integration of
the set of equations (\ref{eq3a}), (\ref{eq3b}) simultaneously gives
the growth factor and logarithmic derivative of the growth factor.
Equations (\ref{eq3a}), (\ref{eq3b}) are valid even when $w$ evolves
with time.

Once the correlation matrix is evaluated, it is straightforward to
obtain expected bounds one can impose on a set of model parameters by
a given data set, thanks to the Fisher information matrix. In linear
regime, the distribution of the counts $n_i$ is multi-variate
Gaussian. In this case, the Fisher information matrix has a simple
form $F_{\alpha\beta} = \frac12 {\rm Tr}({\mbox{\boldmath $C$}}^{-1}
{\mbox{\boldmath $C$}}_{,\alpha} {\mbox{\boldmath $C$}}^{-1}
{\mbox{\boldmath $C$}}_{,\beta})$ \cite{vog96,teg97,mat02}, where
${\mbox{\boldmath $C$}}$ is a theoretical model of the correlation
matrix, $\alpha$, $\beta$ are indices to indicate a kind of model
parameters and ${\mbox{\boldmath $C$}}_{,\alpha}$ is a derivative of
the correlation matrix with respect to a model parameter. An inverse
of a Fisher matrix gives an estimate of the minimum error variance of
a given set of model parameters \cite{the92}.

We consider the LRG sample of the SDSS as a specific example. The
survey area is assumed to be pi steradian, the redshift range to be
0.2--0.4. The number density of the LRGs is approximately homogeneous
and is given by $10^{-4} h^3 {\rm Mpc}^{-3}$ \cite{eis01}. To reduce
the computational cost, we set a sub-region of a $10^2 \pi \simeq 314$
square degree field in a redshift range 0.2--0.4, and fill spherical
cells of radius $15 h^{-1}{\rm Mpc}$ in this region. We assume
nonlinear effects of velocity distortions are erased by this choice of
smoothing radius. In actual applications, it is important to test the
choice of smoothing radius to ensure nonlinear distortions do not bias
the result. We use the cubic closed-packed structure which has the
maximum spatial filling factor of 0.74 that can be filled with
spherical cells without overlapping each other (there are other ways
to achieve this maximum factor such as the hexagonal closed-packed
structure, etc.). As a result, about 2500 cells are placed in this
sub-region. The Fisher information matrix is scaled according to the
ratio of the volume of the sub-region to the total volume.

The model parameters for the dark energy are $\Omega_{\rm Q0}$ and
$w(z)$. To characterize an evolutionary effect of EOS, we employ a
parameterization $w(z) = w_0 + w_1 z$, which is a good approximation
when the evolution of EOS is mild in our interested redshift range of
$z < 0.4$. Therefore, we have three model parameters $\Omega_{\rm
Q0}$, $w_0$, and $w_1$ for the dark energy component. There are other
cosmological parameters to compute the model correlation matrix. We
assume a cold dark matter (CDM) power spectrum which depends on the
shape parameter $\Gamma = \Omega_{\rm M0}h$, and the normalization
$\sigma_8$ of mass. The apparent redshift-space correlation function
depends on $\Omega_{\rm M0}$ and bias parameter $b$, as well as
dark-energy parameters and the power spectrum. In the following we
choose the curvature parameter $\Omega_{\rm K0}$ as an independent
parameter instead of $\Omega_{\rm M0}$. Thus we have seven model
parameters. In the following analysis, we set fiducial values
$\Omega_{\rm Q0} = 0.7$, $w_0 = -1$ (or $-0.5$), $w_1 = 0$,
$\Omega_{\rm K0} = 0$, $h = 0.7$, $\sigma_8 = 1$, $b = 2$, and see how
one can constrain the parameters around this models.

\begin{figure}[ht]
  \includegraphics[height=20pc]{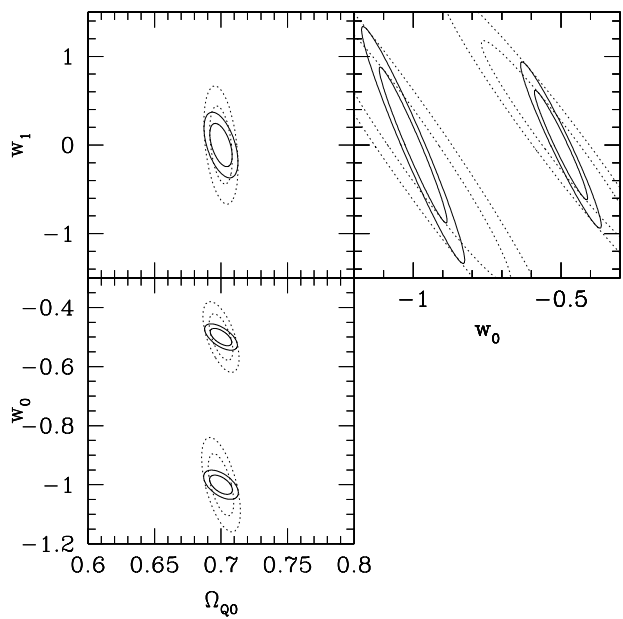}
\caption{Expected error bounds of dark-energy parameters. One of three
parameters are fixed. Inner ellipses represent the $1\sigma$
uncertainty level of one-parameter probability distribution. Outer
ellipses represent $1\sigma$ of the joint probability distribution.
{\em Solid lines}: bias parameter is fixed. {\em Dotted lines}: bias
parameter is marginalized over.
\label{fig1}}
\end{figure}

First, we consider a situation in which only two parameters of the
dark energy are constrained, assuming all the other cosmological
parameters are fixed. Fig.~\ref{fig1} shows the expected error bounds
on two of dark energy parameters are varied.
In $\Omega_{\rm Q0}$-$w_0$ and $w_0$-$w_1$ panels, two cases with
fiducial values $w_0 = -1, -0.5$ are shown, and in $\Omega_{\rm
Q0}$-$w_1$ panel, only $w_0=-1$ case is presented. Inner and outer
ellipses correspond to expected $1\sigma$ bounds of one-parameter
distributions, and $1\sigma$ bounds of two-parameter joint
distributions, respectively. Solid ellipses represent the case in
which bias parameter is fixed, as well as other cosmological
parameters. Since the bias parameter is the most important uncertainty
in analyses of the galaxy clustering, we also plot the expected bounds
marginalized over bias parameter (dotted ellipses). This corresponds
to the case when bias should be simultaneously constrained together
with dark-energy parameters in a same data set.

One may suspect that the constraints on dark energy are imposed mainly
from the evolution of the growth factor $D(z)$, rather than the AP
effect. In which case, the uncertainty in a possible bias evolution
could demolish the constraints. To see this is not the case, we run a
modified code to calculate the Fisher matrix for an imaginary model in
which the growth factor does not evolve at all. The resulting values
of Fisher matrix are only 4\%, 30\%, and 35\% smaller than the
original values for $\Omega_{\rm Q}$-$\Omega_{\rm Q}$, $w_0$-$w_0$,
and $w_1$-$w_1$ components, respectively. Therefore, the AP effect
does mainly constrain the dark energy parameters.

In the $\Omega_{\rm Q0}$-$w_0$ panel, non-evolved EOS models can be
constrained surprisingly well. The bias uncertainty does not
significantly demolish the promise of this aspect. In the $w_0$-$w_1$
panel, the two EOS parameters $w_0$ and $w_1$ are strongly correlated.
This suggests that the redshift range 0.2--0.4 is not large enough to
distinguish evolved model and non-evolved model, and that the
constraints on $w$ mainly depend on an effective value of $w$. In
fact, the degeneracy is represented by $w_0 + 0.13 w_1 = {\rm
const.}$, which means the effective EOS is given by $w_{\rm eff} =
w(z=0.13)$. Since the EOS affects the apparent clustering through
Eq.(\ref{eq1}), it is reasonable that the effective redshift is
halfway to the sample volume.

\begin{figure}[ht]
  \includegraphics[height=20pc]{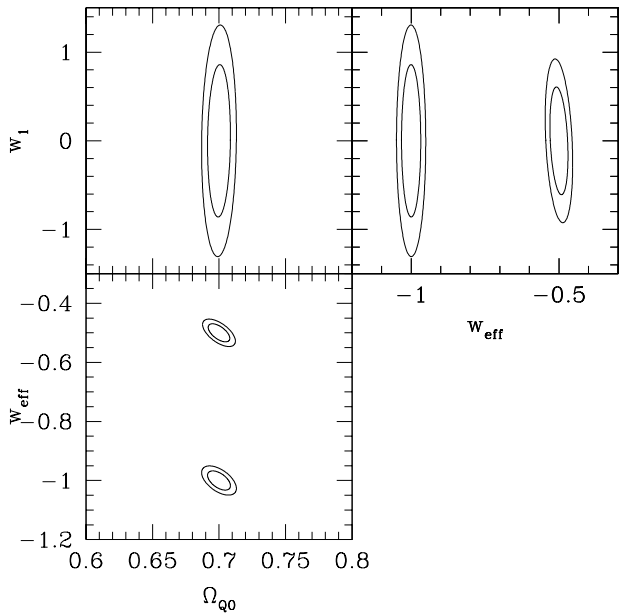}
\caption{Expected error bounds of dark-energy parameters with the
effective EOS $w_{\rm eff} = w(z=0.13)$. One of three parameters are
marginalized over.
\label{fig2}}
\end{figure}

To see how the uncertainty of the EOS evolution affects the
constraints on $\Omega_{\rm Q0}$, $w_{\rm eff}$, we re-parameterize
the EOS by $w(z) = w_{\rm eff} + (z-0.13)w_1$. In Fig.~\ref{fig2},
expected bounds are plotted for the case when one of the dark-energy
parameters $\Omega_{\rm Q0}$, $w_{\rm eff}$, $w_1$ are marginalized
over.
The error bounds on $\Omega_{\rm Q0}$ and $w_{\rm eff}$ are barely
affected by $w_1$ marginalization. This means that our method is not
sensitive to the uncertainty of the EOS evolution, and is only
sensitive to the effective value of the EOS.

\begin{figure}[ht]
  \includegraphics[height=20pc]{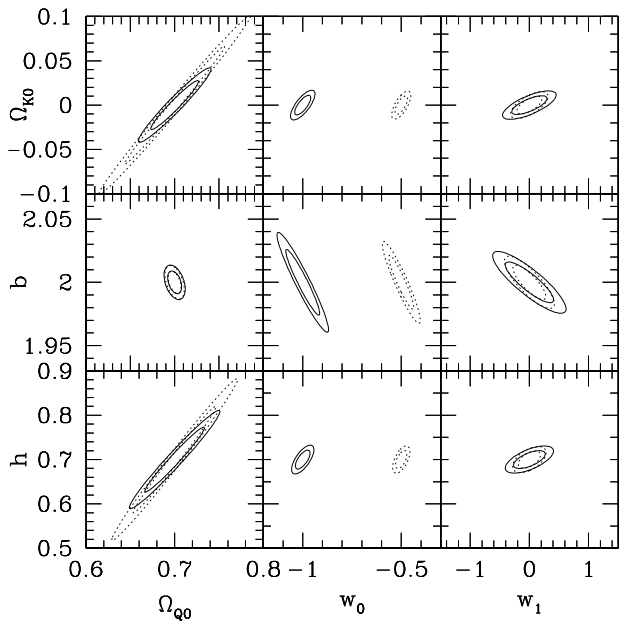}
\caption{Error bounds between dark-energy parameters and other
parameters. {\em Solid line}: $w_0 = -1$ model. {\em Dotted line}:
$w_0 = -0.5$ model.
\label{fig3}}
\end{figure}

In the above analyses, the cosmological parameters other than that of
the dark-energy component and bias parameter are fixed. These other
parameters can be determined from various cosmological observations.
In order to estimate the effects of the uncertainties from these other
parameters, error bounds between dark-energy parameters and other
parameters are plotted in Fig.~\ref{fig3}, without any other
marginalization.
Each panel in this figure indicates a correlation between each pair of
parameters. Correlations with $\sigma_8$ are not plotted, but are
quite similar to those with bias $b$, as naturally expected. The
parameter $\Omega_{\rm Q0}$ is positively correlated with $\Omega_{\rm
K0}$ and $h$. This is because the galaxy clustering accurately
constrains $\Omega_{\rm M0}h = (1 + \Omega_{\rm K0} - \Omega_{\rm
Q0})h$. If we do not use prior knowledge of either the curvature or
the Hubble's constant, the error bound on $\Omega_{\rm Q0}$ is
enlarged by a factor of 4--5. There are negative correlations between
each dark-energy parameter and the bias parameter. This can be
understood by noticing that both parameters increase the amplitude of
fluctuations when the present amplitude is fixed. The EOS parameters
are quite independent on the curvature and Hubble's constant.

In summary, we predict for the first time expected degree of
constraints on dark energy models from intermediate-redshift galaxies
such as LRGs, and show that $\Omega_{\rm Q0}$ and $w_{\rm eff}$ can be
constrained surprisingly well. The prior input of the curvature and
the Hubble's constant are important to reduce the bounds on the dark
energy density, but not so important on the EOS parameters.

We thank A.~Pope and D.~Eisenstein for discussions, and E.~Linder for
comments on the draft. TM acknowledges support from grants MEXT
13740150. AS acknowledges support from grants NSF AST-9802 980 and
NASA LTSA NAG-53503.

\newcommand{\araa}{Annu.~Rev.~Astron.~Astrophys.}
\newcommand{\apjl}{Astrophys.~J.~Lett.}
\newcommand{\aj}{Astron.~J.}
\newcommand{\mnras}{Mon.~Not.~R.~Astron.~Soc.}
\newcommand{\pasp}{Publ.~Astron.~Soc.~Pacific}
\newcommand{\aap}{Astron.~Astrophys.}

\end{document}